\documentclass[12pt,preprint]{aastex}
\usepackage{epsfig}
\usepackage{natbib}
\usepackage{slashbox}
\usepackage{lscape}
\usepackage{mathrsfs,amssymb}
\usepackage{amsmath}
\usepackage{graphicx}	% Including figure files
\usepackage{amssymb}	% Extra maths symbols
\usepackage{subfigure}
\usepackage{ntheorem}
\usepackage{multirow}
\usepackage{array}
\usepackage{booktabs}
\usepackage{makecell}
\usepackage{longtable}
\usepackage{hyperref}
\usepackage{soul}
\usepackage{lineno}
\newcommand       \Angstrom     {\,{\rm \AA}}

\newcommand       \cm           {\,{\rm cm}}

\newcommand       \eV           {\,{\rm eV}}

\newcommand       \K            {\,{\rm K}}

\newcommand       \simlt        {\lesssim}
\newcommand       \simgt        {\gtrsim}

\newcommand       \mum          {\,{\rm \mu m}}
\newcommand       \ppm          {\,{\rm ppm}}

\newcommand       \simali       {\sim\,}

\newcommand       \km        {\,{\rm km}}
\newcommand       \Liter      {\,{\rm L}}
\newcommand       \mol       {\,{\rm mol}}
\newcommand       \mole       {\,{\rm molecule}}

\newcommand       \varepsilonWV {\varepsilon_{\tilde{\nu}}}
\newcommand       \varepsilonmax {\varepsilon_{\tilde{\nu},{\rm max}}}

\newcommand	  \ssun        {\left[{\rm S/H}\right]_{\odot}}

\newcommand	  \sism        {\left[{\rm S/H}\right]_{\scriptsize\rm ISM}}

\newcommand	  \sdust        {\left[{\rm S/H}\right]_{\scriptsize\rm dust}}

\newcommand	  \sgas        {\left[{\rm S/H}\right]_{\scriptsize\rm gas}}

\newcommand	  \fegas       {\left[{\rm Fe/H}\right]_{\scriptsize\rm gas}}

\newcommand \cpah       {\left[{\rm C/H}\right]_{\scriptsize\rm PAH}}
\newcommand \spah       {\left[{\rm S/H}\right]_{\scriptsize\rm PAH}}
\newcommand	  \NC        {N_{\rm C}}
\newcommand	  \NS       {N_{\rm S}}
\countdef\decade=200
\advance\decade by \year
\countdef\hours=201
\advance\hours by \time
\divide\hours by 60
\countdef\mins=202
\advance\mins by \hours
\multiply\mins by 60
\multiply\hours by 100
\countdef\miltime=203
\advance\miltime by \hours
\advance\miltime by \time
\advance\miltime by -\mins
\def\today{\number\decade.\number\month.\number\day.\number\miltime}
\shorttitle{Where Have All the Sulfur Atoms Gone?}
\title{
Where Have All the Sulfur Atoms Gone?
Polycyclic Aromatic Hydrocarbon
as a Possible Sink for the Missing Sulfur
in the Interstellar Medium.
I. The C--S Band Strengths
\\{\small DRAFT: \today ~~}
}
\author{
X.J.~Yang\altaffilmark{1,2},
Lijun~Hua\altaffilmark{1}, and
Aigen Li\altaffilmark{2}
}
\altaffiltext{1}{Hunan Key Laboratory for Stellar and Interstellar Physics and School of Physics and Optoelectronics, Xiangtan University, Hunan 411105, China
                      {\sf xjyang@xtu.edu.cn}}
\altaffiltext{2} {Department of Physics and Astronomy,
                  University of Missouri,
                  Columbia, MO 65211, USA;
                  {\sf lia@missouri.edu}}

\begin{document}
%\linenumbers
\begin{abstract}
%{\bf
Despite its biogeneic and astrochemical importance,
sulfur (S), the 10th most abundant element
in the interstellar medium (ISM)
with a total abundance of
S/H\,$\approx$\,$2.2\times10^{-5}$,
largely remains undetected in molecular clouds.
%}
%S-containing molecules and ices
%fall short of accounting for the interstellar S abundance
%of $\simali$$2.2\times10^{-5}$ (relative to H)
%by a factor of $\simali$1000, indicating that the majority
%of the missing S in molecular clouds
%must have been depleted in dust.
Even in the diffuse ISM where S was previously often
believed to be fully in the gas phase, in recent years
observational evidence has suggested that S may also
be appreciably depleted from the gas.
What might be the dominant S reservoir in the ISM
remains unknown. Solid sulfides like MgS, FeS and
SiS$_2$ are excluded as a major S reservoir
due to the undetection of their expected infrared
spectral bands in the ISM.
In this work, we explore the potential role of
sulfurated polycyclic aromatic hydrocarbon (PAH)
molecules---PAHs with sulfur heterocycles
(PASHs)---as a sink for the missing S. Utilizing
density function theory, we compute the vibrational
spectra of 18 representative PASH molecules.
It is found that these molecules exhibit a prominent,
C--S stretching band at $\simali$10$\mum$ and
two relatively weak, C--S deformation bands
at 15 and 25$\mum$ that are not mixed with
the nominal PAH bands at 6.2, 7.7, 8.6, 11.3
and 12.7$\mum$.
If several parts per million of S (relative to H)
are locked up in PAHs, the 10$\mum$ C--S band
would be detectable by {\it Spitzer} and JWST.
%{\bf
To quantitatively explore the amount of S/H
depleted in PASHs, detailed comparison of
the infrared {\it emission} spectra of PASHs
with the {\it Spitzer} and JWST observations
is needed.
%}
%
\end{abstract}

\section{Introduction\label{sec:intro}}
Sulfur (S) is the 10th most abundant element in
the interstellar medium (ISM).
As a biogenic element, S plays a crucial role in the
biological systems on Earth (Krijt et al.\ 2023 and references therein).
In the diffuse ISM where there are plenty of
ultraviolet (UV) photons
with energies exceeding 10.36$\eV$,
the ionization potential energy of S,
the predominant form of gas-phase S
in the diffuse ISM is sulfur ion (S$^{+}$).
In molecular clouds, S is found in gas molecules
and ices (e.g., see Mifsud et al.\ 2021,
Hily-Blant et al.\ 2022, McClure et al.\ 2023).
The abundance of S relative to hydrogen (H),
S/H, is an essential parameter for understanding
the interstellar gas chemistry.
S-bearing molecules are often used as tracers
of their kinematics and chemical evolution
in star- and planet-forming clouds
\citep{Zhou1993,Wakelam2004,Dutrey1997,LeGal2019}.
%(in reference list of Cazaux et al. 2022, A&A, 657, A100)
%
However, our understanding of the interstellar
S/H abundance is still poor or at least incomplete:
it remains notoriously mysterious
regarding the possible carriers of the S elements
missing from the gas phase in molecular clouds;
it remains unknown whether S in the diffuse ISM
is partly depleted from the gas and locked up in dust;
even more fundamentally, it is not definitely clear
what the true interstellar S/H abundance is.

\subsection{What is the interstellar S/H abundance?
                     \label{subsec:Sabundance}}
Elements in the ISM exist in the form of gas or dust.
The interstellar gas-phase abundance of an element
can be measured from its optical and UV absorption lines.
The elements ``missing'' from the gas phase are bound up
in dust grains, known as ``interstellar depletion''.
The total interstellar abundance of an element
is the sum of its gas-phase abundance and
the interstellar depletion.
The interstellar depletion of an element is commonly
derived from the interstellar extincion and infrared (IR)
emission modeling and/or the UV
(e.g., the 2175$\Angstrom$ extinction bump)
or IR spectral bands
(e.g., the 9.7$\mum$ silicate absorption band).
This requires an assumption of dust size and composition
which are often not accurately known.

Therefore, instead of summing up the gas-phase
abundance and the interstellar depletion, one often
assumes a ``reference standard'' for the interstellar abundance.
Historically, the Sun is such a common reference standard,
i.e., the interstellar abundances of elements are assumed
to be those of solar measured from the solar photosphere.
In this case, the interstellar S/H abundance would be
$\sism=\ssun\approx14.5\pm1.0\ppm$ \citep{Asplund2009}.
Because of their young ages,
it had also been suggested that B stars
and young F, G stars might better represent
the interstellar abundances.
While the carbon abundance (C/H) of B and
young F, G stars are appreciably lower than solar,
their S/H abundance of $\simali$$16.2\pm1.3\ppm$
is similar to solar (within uncertainties)
or somewhat higher than solar (see Przybilla et al.\ 2008).
%
%It has been argued that the present-day solar
%photospheric abundance (as well as that of B stars
%and young F, G stars) may be lower than
%the present-day interstellar abundance.
Lodders (2003) pointed out that
the present-day solar photospheric abundances
(as well as that of B stars and young F, G stars)
could be lower than those of the proto-Sun
because heavy elements might have settled
toward the Sun's interior since the time of
its formation $\simali$4.55\,Gyr ago.
Taking the settling effects into account,
Lodders (2003) derived a protosolar
S/H abundance of $18.2\pm1.7\ppm$.
Compared to the ISM out of which the Sun was formed
4.55\,Gyr ago, the present-day ISM has been chemically
enriched over the past 4.55\,Gyr
due to the Galactic chemical evolution (GCE).
Assuming a GCE enrichment of
0.09 dex \citep{Chiappini2003},
we obtain a {\it recommended},
present-day interstellar S/H
abundance of $22.4\pm{1.7}\ppm$.
%In the following, we will adopt
%$\sism=22.4\ppm$.

\subsection{Is S Depleted in the Diffuse ISM?
                     \label{subsec:Sdepletion}}
In the diffuse ISM, it was once believed that S is
entirely volatile and there is no S depletion in solids.
However, Howk et al.\ (2006) measured the gas-phase
S/H abundance for the Galactic halo and derived
$\sgas\approx 13.5\ppm$. While close to the solar
abundance of $\ssun\approx 14.5\ppm$ \citep{Asplund2009},
the Galactic halo gas-phase S/H abundance
is considerably lower than the interstellar abundance
of $\sism=22.4\ppm$
as discussed in \S\ref{subsec:Sabundance},
implying that S is partly depleted in dust
even in the Galactic halo.
Howk et al.\ (2006) also measured
the Galactic halo gas-phase abundance
of iron (Fe) to be $\fegas\approx 5.4\ppm$.
Although in the Galactic halo Fe is not as heavily
depleted as in the diffuse ISM,
the Galactic halo gas-phase Fe/H abundance
is still substantially lower than the interstellar
Fe/H abundance of $\simali$48$\ppm$
(see Zuo et al.\ 2021),
indicating a significant incorporation of Fe into dust.
This demonstrates that dust grains are present
and survive in the hostile halo environment.
Therefore, it is not unreasonable to have some
portions of the interstellar S elements
to be locked up in dust in the Galactic halo.

\citet{White2011} analyzed the strong
S\,II 1250, 1253, 1259$\Angstrom$ absorption bands
toward 28 interstellar lines of sight
obtained with the high-resolution
Space Telescope Imaging Spectrograph (STIS) instrument
on board the Hubble Space Telescope (HST).
Their results were consistent with the depletion
of S into dust grains, although it does not follow
the pattern of other elements in dust.
\cite{Jenkins2009} proposed a unified presentation
of elemental depletion in the ISM and also concluded
that S is depleted from the gas phase in the diffuse ISM.
\cite{Hensley2021} adopted $\sgas\approx 9.6\ppm$
and this would suggest a depletion of
$\sdust\approx 12.8\ppm$ for the diffuse ISM,
if we adopt an interstellar abundance of
$\sism=22.4\ppm$.

\subsection{Extreme Depletion of S in Dense Clouds
                    \label{subsec:denseclouds}}
The S/H abundance in the dense ISM is a long-standing
problem in astrochemistry (e.g., see Laas \&
Caselli 2019 and references therein).
Many observations have shown that S depletion
is far more severe in dense clouds than in the diffuse ISM.
%The abundances of S-containing molecules and ions
%for dense molecular clouds are lower than that expected
%from the interstellar S/H abundance by a factor of
%as much as a thousand
%(Prasad \& Huntress 1982; Anderson et al. 2013).
The observed abundances of S-bearing molecules
in molecular clouds (e.g., CS, H$_2$S, SO, SO$_2$ and OCS)
make up a very small fraction of the S nuclei
(Prasad \& Huntress 1982; Anderson et al. 2013).
For example, in cold dark clouds and dense cores
shielded from stellar UV radiation
where most S is expected to be in molecular form,
by adding the abundances of all detected
gas-phase S-bearing molecules,
the resulting total S/H abundance is typically
a factor of $\simali$10$^2$--10$^3$ lower
than the interstellar S/H abundance
($<$\,1$\ppm$; e.g., see Fuente et al.\ 2019).
It is not yet clear in which form most S resides
in molecular clouds.
The S depletion appears to be environment-dependent
\citep{Fuente2023}.\footnote{%
%  {\bf
    By analyzing the atomic and ionized lines
    of S detected by the James Webb Space Telescope
    (JWST), Fuente et al.\ (2024) derived
    $\sgas\approx 8\ppm$
    in the ionized and warm molecular phases
    toward the Orion Bar.
    Daflon et al.\ (2009) estimated
    the sulfur abundance based on
    the photospheric lines of a sample of ten
    B main-sequence stars of the Orion association
    and obtained S/H\,$\approx$\,$14.1\pm1.7\ppm$.
    %which is consistent with the solar value
    %and that of meteorites (Asplund et al.\ 2006).
    %Esteban et al.\ (2004) observed a region near
    %the hot star ${\rm \Theta}^1$ Ori, and measured
    %$\sgas\approx16.5\ppm$.
   This suggests a depletion of
   $\sdust\approx 6\ppm$ in the Orion Bar,
   implying a lower depletion of S/H in regions
   illuminated by young stars.
   }
%   }
%
%{\bf
More recently, Ferrari et al.\ (2024) suggested
that octasulfur S$_8$ may be an important S reservoir
in molecular clouds, formed upon UV irradiation of
H$_2$S ices or electron irradiation of H$_2$S and
SO$_2$ ices (e.g., see Shingledecker et al.\ 2020,
Mifsud et al.\ 2021, Cazaux et al.\ 2022).
It is interesting to note that S$_8$ was recently
detected in the Ryugu asteroidal samples
(Aponte et al.\ 2023). If S$_8$ is responsible for
the missing S in molecular clouds, it is expected
to reveal its presence through three prominent
bands at 53.5, 41.3 and 21.1$\mum$
(Ferrari et al.\ 2024).
%}

\subsection{Where Have All the Missing S Atoms Gone?
                    \label{subsec:Sdepletion_carrier}}
In the ISM, elements not seen in the gas phase
must have been locked up in dust.
For example, at least 95\% of the interstellar Si,
Mg and Fe elements are missing from the gas phase,
indicating Si and Mg (and probably Fe as well)
are locked up in silicate dust.
The case for Fe is more complicated.
The missing Fe could be depleted in Fe-rich silicate dust,
iron oxides, iron sulfides (FeS) or pure solid iron.
For the S elements missing from the gas phase,
popular candidate solids are sulfides,
including magnesium sulfide (MgS),
silicon disulfide (SiS$_2$) and, as mentioned above, FeS.

The formation of sulfide is likely to occur
in carbon-rich circumstellar environments.
Lattimer et al.\ (1978) predicted the possible
presence of various sulfur-bearing materials
in carbon-rich systems.
According to Zhukovska et al.\ (2008),
%\citet{Zhukovska2008A&A...486..229Z},
MgS has priority in the cooling sequence of
sulphur-bearing solid compounds.
In principle, upon injection into the ISM,
sulfides could be an component of interstellar dust.

However, MgS exhibits a prominent band
at 30$\mum$. While such a 30$\mum$
emission band is seen in
asymptotic giant branch (AGB) stars,
post-AGB stars and planetary nebulae,
it has never been seen in the ISM.
The 30$\mum$ band is neither seen
in absorption in heavily obscured
interstellar sources toward the Galactic center
nor in Wolf-Rayet stars.
This suggests that MgS is not
an interstellar dust component
and the majority of the missing S
is not depleted in MgS.

Similarly, the laboratory spectrum of SiS$_2$
displays a prominent feature at $\simali$22$\mum$
and a secondary feature at 17$\mum$
%\citep{Nuth1985ApJ...290L}.
(Nuth et al. 1985).
As neither the 22$\mum$ feature
nor the 17$\mum$ feature is seen
in the ISM, either in absorption or emission,
SiS$_2$ cannot be an interstellar dust component
and, thus SiS$_2$ cannot be the sink of the missing S.

FeS has been identified in cometary dust sample
and primitive meteorites
 (e.g., see Zolensky et al.\ 2006).
Submicron-sized GEMS
(glass with embedded metals and sulfides)
have been identified in interplanetary dust particles
(IDPs, Bradley 2003).
Laboratory spectra of FeS solids show a strong band
centered at $\simali$23.5$\mum$.
While this band has been detected in emission
in circumstellar disks
surrounding several young stars
and protoplanetary nebulae (PPNe)
around evolved stars \citep{Keller2002},
it has not been seen in the ISM,
either in emission or absorption.
Therefore, FeS is also excluded as
an important interstellar dust component.

In dense clouds, a possible explanation
for the missing S is that,
in addition to octasulfur S$_8$
(see Shingledecker et al.\ 2020,
Ferrari et al.\ 2024), S may be present
in the icy mantles that coat interstellar
dust grains.\footnote{%
%  {\bf
  Ammonium hydrosulfide (NH$_4$SH) salt
  may also be present in dense clouds and
  in the upper layer of the coma in comets
  such as 67P/Churyumov-Gerasimenko
  (Altwegg et al.\ 2020).
  Very recently, Vitorino et al.\ (2024) performed
  experimental studies and demonstrated that
  NH$_4$SH salt could be formed at $\simali$10$\K$
  on grains from a mixture of H$_2$S and NH$_3$.
  Santos et al.\ (2024) also showed experimentally
  that S-bearing complex organic molecules could
  be formed from the interaction between C$_2$H$_2$
  molecules and SH radicals in interstellar ice mantles
  at $\simali$10$\K$.
  }
%  }
However, ice carriers of
S reservoir including solid OCS, H$_2$S,
and SO$_{2}$ can only account for a tiny
fraction of the missing S
 (e.g., see Laas \& Caselli 2019
and references therein).
% \citep{Laas2019,Shingledecker2020,Cazaux2022}.
%
In Table~\ref{tab:Sabundance} we summarize
the interstellar S/H abundance as approximated
by the photospheric abundance of the Sun,
B stars and young F and G stars,
proto-Sun, and the protosolar abundance
argumented by the GCE.
Also tabulated is the S/H abudnace missing
from the gas phase in the diffuse ISM
and in molecular clouds.

\subsection{PAH as a Possible Sink for the Missing S
                   \label{subsec:Sdepletion_PASH}}
In \S\ref{subsec:Sdepletion_carrier} we have discussed
that sulfides are unlikely the sink for the missing S
in the ISM. Then, what else could the S reservior be?
As the smallest carbon component of the interstellar
dust populations, polycyclic aromatic hydrocarbon (PAH)
molecules are abundant and widespread in the Universe,
as revealed by the ubiquitous detection of
the aromatic IR emission (AIE) bands at 3.3,
6.2, 7.7, 8.6, 11.3 and 12.7$\mum$
\citep{Leger1984,Allamandola1985,Tielens2008,Li2020}.
Astronomical PAHs may not be pure aromatic compounds
as strictly defined by chemists.
Instead, PAH molecules in astronomical
environments may include ring defects,
aliphatic component (e.g., aliphatic sidegroups
like methyl --CH$_3$; see Yang \& Li 2023a),
substituents (e.g., N in place of C,
see Hudgins et al.\ 2005, Mattioda et al.\ 2008;
O in place of C, see Bauschlicher 1998;
Fe in place of C, see Szczepanski et al.\ 2006),
partial deuteration
(e.g., see Allamandola et al.\ 1989,
Draine 2006, Yang \& Li 2023b),
partial dehydrogenation
and sometimes superhydrogenation
(e.g., see Bernstein et al.\ 1996,
Sandford et al.\ 2013, Yang et al.\ 2020).

\begin{figure*}
 \vspace{-2mm}
  \begin{center}
  \epsfig{file=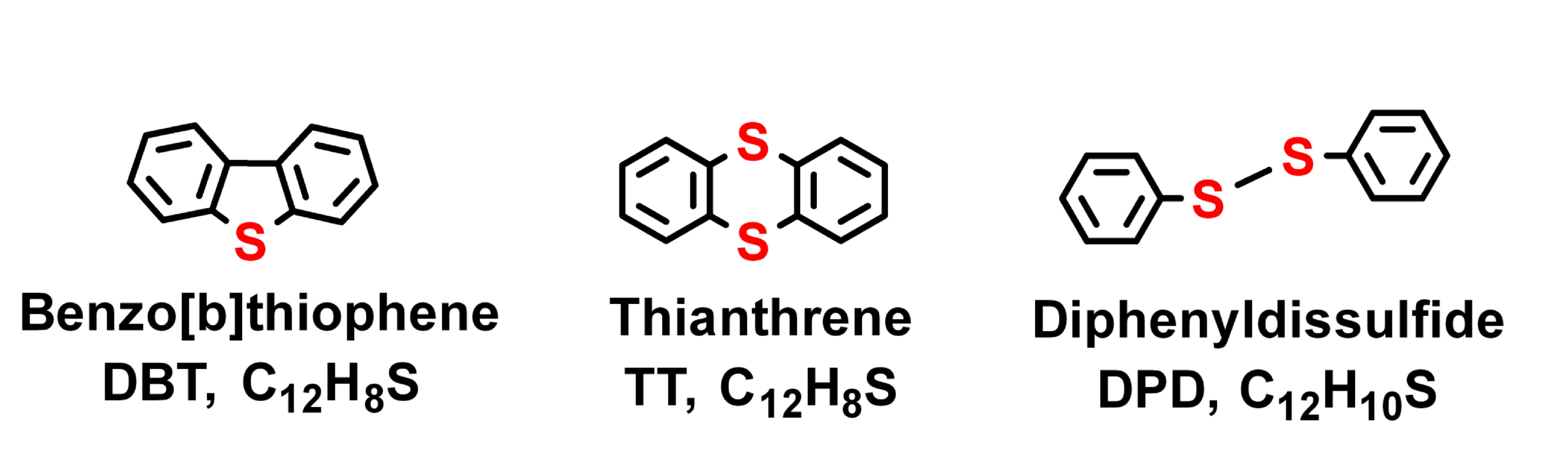,width=12cm}
  \end{center}
\vspace{-1mm}
\caption{
\label{fig:MeteoritePASH}
 Structures of PASH molecules identified in meteorites.
        }
\vspace{-3mm}
\end{figure*}

Astronomical PAHs may also be sulfurated.
%\citep{OrthousDaunay2010}
Orthous-Daunay et al.\ (2010) performed
XANES (X-ray Absorption Near Edge Structure)
spectroscopic measurements of
carbonaceous chondrites and identified
several polycyclic aromatic sulfur heterocycles
(PASHs; see Figure~\ref{fig:MeteoritePASH}).
If PASHs are present in the ISM in an appreciable
quantity, it may not be impossible that PASHs
may (at least partly) account for the missing S mystery.
This motivates us to explore the IR emission spectra
of interstellar PASHs.

To facilitate future searches for
PASH molecules and quantitative determination
of the PASH abundance so as to evaluate the potential
of PASHs as a reservior of the missing S in the ISM,
especially in the JWST era,
we perform for the first time a systematic calculation
of the IR vibrational spectra of various PASH species.
The structure of this paper is as follows.
In \S\ref{sec:methods} we briefly describe
the computational methods and target molecules.
 We show in \S\ref{sec:results} the calculational
results and derive the mean spectral properties
(particularly the C--S band strengths)
of PASH molecules (see \S\ref{subsec:CS_spectra}).
We then implement them in the astro-PAH model
of Li \& Draine (2001) and Draine \& Li (2007)
and discuss the astrophysical implications
in \S\ref{sec:discussion}.
Finally, we summarize our major results
in \S\ref{sec:summary}.
This paper is largely concerned
with the C--S band strengths of PASHs.
In a subsequent paper we will model the IR
emission spectra of PASH molecules of various
sizes and compare with the astronomical spectra
obtained with the Spitzer Space Telescope and JWST
to quantitatively derive the amount of S/H
depleted in PASHs.

\section{Computational Methods
            and Target Molecules\label{sec:methods}}
\begin{figure*}
 \vspace{-2mm}
  \begin{center}
  \epsfig{file=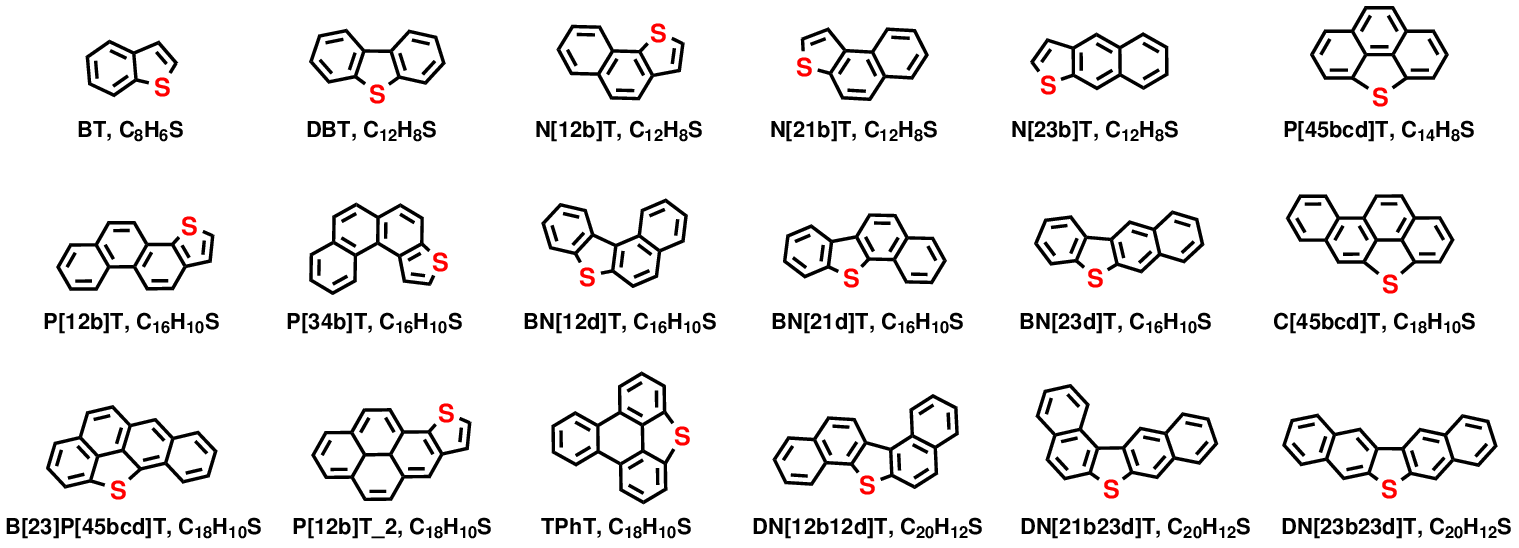,width=18cm}
  \end{center}
\vspace{-1mm}
\caption{\label{fig:TargetMolecules}
      Structures of all the target molecules considered in this work.
        }
\vspace{-3mm}
\end{figure*}

PASHs have been widely studied by chemists,
since the presence of heteroatoms of sulfur
allows for diversifying the structures,
reactivity and electronic properties
 (see Delaunay et al.\ 2016 and references therein).
%\citep[and reference \textbf{therein}]{Delaunay2016}.
%
However, there is little study on PASHs in astronomy.
As an illustrative investigation,
in this work we consider target PASH molecules
as shown in Figure~\ref{fig:TargetMolecules}.
We focus on the heterocyclic organic sulfur,
which has one or more benzene rings
and also a five-membered ring with S incorporated.
We consider PASH molecules
with only one S atom incorporated.
Those with more than one S atoms
will be considered later.
We start from one benzene ring
all the way up to five.\footnote{%
    We note that these are fairly small molecules.
    Molecules with more than 20 C atoms
    (up to several hundreds and thousands
    of C atoms) are expected to emit in the JWST
    wavelength regime (see Croiset et al.\ 2016,
    Maragkoudakis et al.\ 2020, Draine et al.\ 2021).
    As we are mostly interested in the C--S band
    strengths of PASHs, it is not unreasonable
    to focus on small molecules. Also, large molecules
    are computationally expensive
    (and often not practical).
    }
Furthermore, we consider several isomers
with different configurations.
%
%We do not go to larger molecules
%since \textbf{they have larger heat capacity,
%therefore upon absorbing a photon
%they will not be heated to the
%temperature which is required to }
%emit efficiently in the near-to-mid IR
%wavelength range covered by JWST.
%
In total, we consider 18 molecules and their cations.
The abbreviation of the naming of these molecules
are referred following the National Institute of Standards
and Technology (NIST) webbook.\footnote{%
  https://webbook.nist.gov/chemistry/
  }

We use the Gaussian16 software \citep{Frisch2016}
to calculate the IR spectra of all of our target molecules.
We employ the hybrid density functional theoretical
(DFT) method (B3LYP) at the {\rm 6-311+G$^{\ast\ast}$}
level to achieve the best compromise
between accuracy and computerational time
(see Yang et al.\ 2013).
The calculated frequencies are scaled with
a factor of 0.9688 for all the vibrational modes
\citep{Borowski2012}.
We note that in the literature sometimes
different scaling factors were used for different
modes (e.g., see Bauschlicher et al.\ 2018).

\section{Results\label{sec:results}}
\subsection{Computational Accuracy
                    \label{sec:accuracy}}
To test the accuracy of our calculation,
we first compare the calculated IR spectra
with those measured from laboratory.
We apply this accuracy test
to three molecules (BT, DBT and TT;
see Figure~\ref{fig:TargetMolecules}
for molecule naming abbreviation)
since experimental spectra are available
for these three molecules.

\begin{figure*}
 \vspace{-2mm}
  \begin{center}
  \epsfig{file=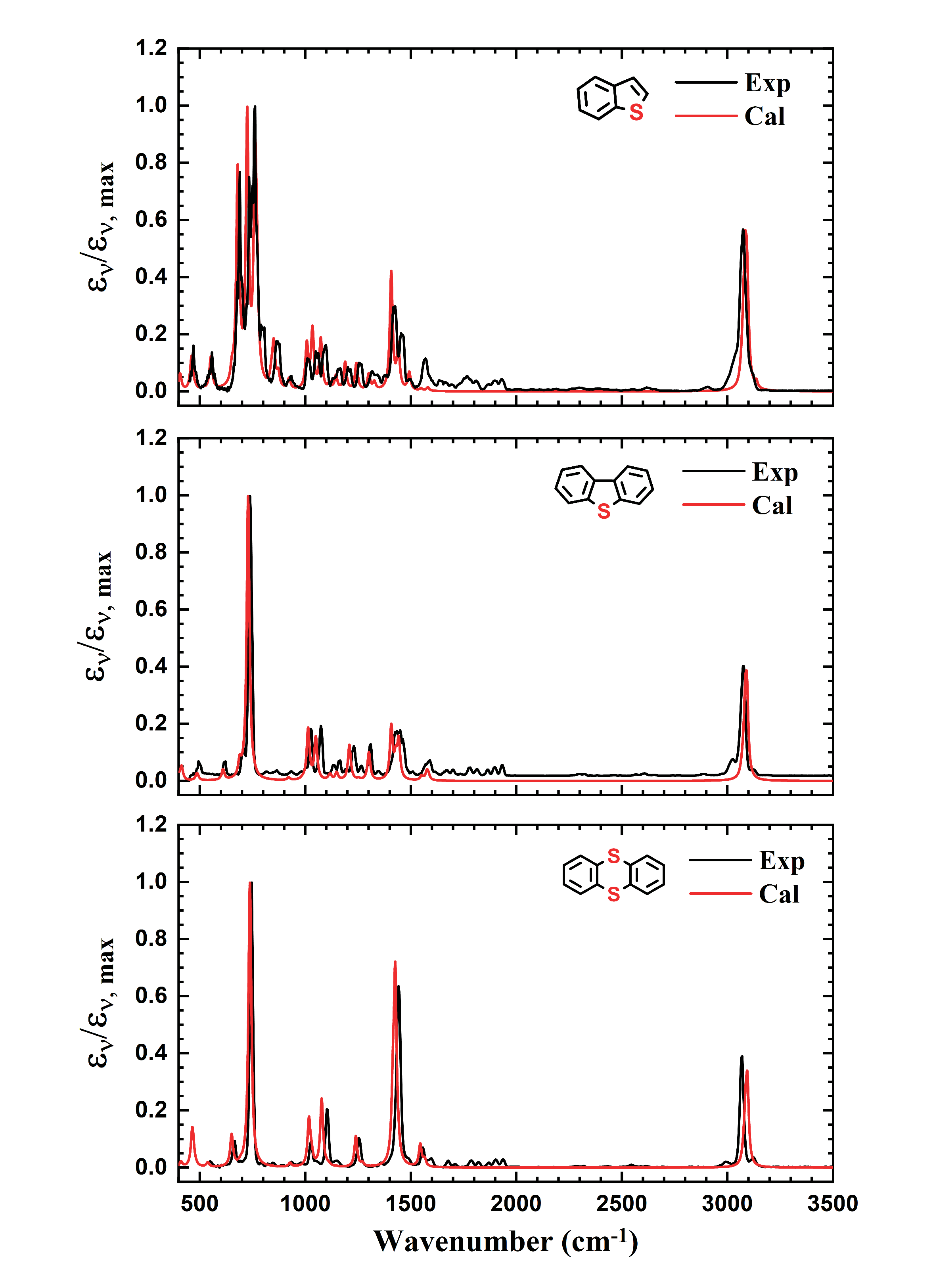,width=12cm}
  \end{center}
\vspace{-1mm}
\caption{\label{fig:PASH_Exp_Cal}
         Comparison of the experimental (black solid lines) and
         calculational (red solid lines) spectra for BT (upper panel),
         DBT (middle) and TT (bottom). The experimental spectra are
         taken from the NIST webbook.
         The frequencies of the calculated spectra are scaled with
         a factor 0.9688, and an
         FWHM of 16~$\cm^{-1}$ is assigned.
         For comparison, both the experimental and calculated spectra
         are normalized to their maxima
         due to the lack of experimental information
         on the intensities of these molecules.
         }
\vspace{-3mm}
\end{figure*}

Figure~\ref{fig:PASH_Exp_Cal} compares
the computational spectra of BT, DBT and TT
with their experimental spectra
taken from the NIST webbook.
As we do not have experimental information
for the band intensities,
we focus on the relative band strength comparison.
The spectra are expressed as the molar (mol)
extinction coefficient $\varepsilonWV$
normalized to its peak value $\varepsilonmax$,
where $\tilde{\nu}\equiv\lambda^{-1}$
is the wavenumber.
The full width at half maximum (FWHM)
for all the calculational spectral bands
is set to be 16$\cm^{-1}$ to match
the experimental spectra.

Figure~\ref{fig:PASH_Exp_Cal} shows that
we basically achieve consistency between
calculational spectra and experimental spectra
by employing the current theoretical level
with the scaling factor given in Gaussian16.
The peak wavelengths for the main vibrational modes,
such as the C--H stretch at $\simali$3100$\cm^{-1}$,
the ${\rm C=C}$ strtech at $\simali$1400$\cm^{-1}$, and
the C--H out-of-plane bending at $\simali$750$\cm^{-1}$,
match very well between calculations and experiments.
A few minor features shown in the experimental spectra
at $\tilde{\nu}$\,$\simali$1500--2000$\cm^{-1}$
are absent in the calculational spectra,
suggesting that they are not fundamentals but
probably overtones or combinations instead.

Figure~\ref{fig:PASH_Exp_Cal} demonstrates that
our computational method can reasonably
well reproduce the vibrational spectra of PASH molecules,
both in wavelengths and in relative band strengths.
In this regard, we will apply the computational method
to all our 18 target molecules
with B3LYP at level {\rm 6-311+G$^{\ast\ast}$}.

\subsection{BT and its Cation\label{subsec:BT_spectra}}
\begin{figure*}
 \hspace{-2mm}
  \begin{center}
  \epsfig{file=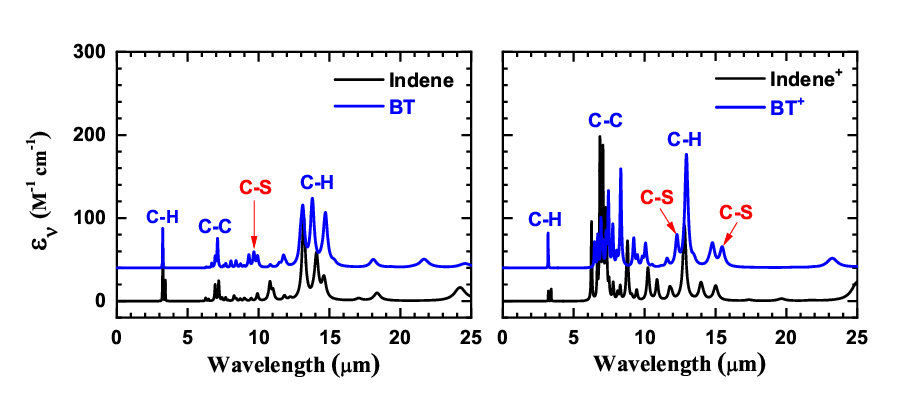,width=18cm}
  \end{center}
\vspace{-10mm}
\caption{\label{fig:Spec_1BenzoRing}
 Calculated spectra of BT (left panel) and BT$^{+}$ (right panel).
 The origins of the main features are labeled, including
 C--H, C--C, and C--S bonds.
 The frequencies of the calculated spectra are scaled with
 a factor 0.9688, and a FWHM of 16$\cm^{-1}$ is assigned.
 }
\vspace{-3mm}
\end{figure*}

We show in Figure~\ref{fig:Spec_1BenzoRing} the calculated
IR spectra of BT and its cation. To illustrate the vibrational
bands more clearly, we adopt a FWHM of 16$\cm^{-1}$.
We clearly see in Figure~\ref{fig:Spec_1BenzoRing}
the C--H stretching band at 3.3$\mum$,
C--C stretch and C--H in-plane (ip)
bending bands at $\simali$6--10$\mum$,
as well as the C--H out-of-plane (oop)
bending bands at $\simali$10--15$\mum$.
For pure PAHs, it is well known that neutrals
have stronger 3.3 and 11.3$\mum$ C--H bands,
while cations contribute more to the C--C stretching
and C--H ip bending bands in the wavelength
range of $\simali$6--10$\mum$
(e.g., see Allamandola et al.\ 1999).
However, for the PASH molecule BT,
the situation seems different.
While the BT cation (BT$^{+}$)
has much strong C--C stretches than neutral BT,
the 3.3$\mum$ C--H stretches are equally strong
for BT ($\simali$7.34$\km\mol^{-1}$)
and its ionic counterpart BT$^{+}$
($\simali$5.84$\km\mol^{-1}$).
Also, the C--H oop  bending bands
at $\simali$10--15$\mum$ for BT
is even weaker than that of BT$^{+}$.

To examine if such an effect is caused
by sulfuration, we also show
in Figure~\ref{fig:Spec_1BenzoRing}
the IR spectra of neutral and cationic indene,
the {\it pure} PAH version (i.e., no sulfuration)
of BT and BT$^{+}$. It is apparent that, just like
typical PAH molecules, both the C--H stretch
at 3.3$\mum$ and the C--H oop bending bands
at $\simali$10--15$\mum$ of neutral indene
are much stronger than that of indene cation.
This indicates that sulfuration significantly affects
the C--H vibrational intensities of BT and BT$^{+}$.

Figure~\ref{fig:Spec_1BenzoRing}
also shows that sulfuration also results in wavelength
shifts for some of the vibrational bands.
Generally, the incorporation of a metal element
in PAHs will cause the shift of the central wavelength
of the PAH bands, especially for the 6.2$\mum$ C--C
stretch. As a matter of fact, N-incorporated PAHs
are usually resorted to explain the different central
wavelength for the observed 6.2$\mum$ emission band,
and it is suggested that the N atom will cause the variation
of the distribution of the electronic cloud
(e.g., see Hudgins et al.\ 2004, Ricca et al.\ 2021).
Experimental and computational studies have shown
that the ``6.2$\mum$'' band of {\it pure} PAHs occurs
at an appreciably longer wavelength incomparable
with the astronomical 6.2$\mum$ PAH band.\footnote{%
    However, Ricca et al.\ (2024) recently argued
    that PAHs with edge defects could explain the central
    wavelength of the astronomical 6.2$\mum$ PAH band
    using spectra computed with DFT without introducing
    PAHs that contain nitrogen within their ring structures.
    }
%    }
%

The C--S bonds give rise to stretching vibrations,
defined by the significant changes of the bond lengths.
They generally occur in the $\simali$9--16$\mum$
wavelength range. Meanwhile, there are also deformation
vibrations, where the bond angles change,
lying in the wavelength range of $\simali$15--22$\mum$.
For illustration, we label the C--S stretch and deformation
vibrations as ``C--S'' in Figure~\ref{fig:Spec_1BenzoRing}.
%
%{\bf
As we are primarily interested in the C--S
band strengths of PASHs, in the following we will
focus on the C--S vibrations.
For the C--C and C--H band strengths
for ``model'' PASH molecules,
we will {\it not} adopt that of the PASH target
molecules computed here, instead, we will
adopt that of astro-PAHs (Li \& Draine 2001,
Draine \& Li 2007, Draine et al.\ 2021;
see \S\ref{sec:discussion}).
The underlying assumption of this is that
we assume that the presence of S atoms in
a PAH molecule would not affect the C--C
and C--H vibrations.
%}

\subsection{C---S Vibrational Spectra
                    \label{subsec:CS_spectra}}
We identify the C--S vibrations
from the Gaussian output file,
and read their frequencies and intensities
for all the calculated target molecules and their cations.
These results are tabulated in
Tables~\ref{tab:CS_Freq_Int_Neutral}
and \ref{tab:CS_Freq_Int_Cation}
for the neutrals and cations, respectively.

The C--S stretches basically fall into
the wavelength range of $\simali$9--16$\mum$,
while the C--S deformation bands occur mainly
in $\simali$15--22$\mum$ except several cations
show deformation bands at $\simali$12$\mum$.
All the molecules considered here show relatively
strong C--S stretch around $\simali$10$\mum$
and $\simali$12$\mum$, and several satellite features
at $\simali$10$-$15~$\mum$.
The deformation vibrations often peak at
$\simali$15$\mum$ and $\simali$20$\mum$.
Generally, for both neutrals and cations
the stretching modes have much larger
intensities than the deformation modes.
Both vibrational modes show a large diversity
of wavelengths and intensities.
The vibration modes often couple with each other.
Especially in our target molecules,
the five-membered ring where the S atom locates
is attached to one or two benzene rings.
Since several carbon atoms are shared
by the  five- and six-membered rings,
the vibrations of these involved carbon atoms
are often coupled.
The degree of coupling is dependent on the configuration.
Therefore, for different molecules, the number of
C--S vibrational modes is different.
%

\begin{figure*}
 \hspace{-2mm}
  \begin{center}
  \epsfig{file=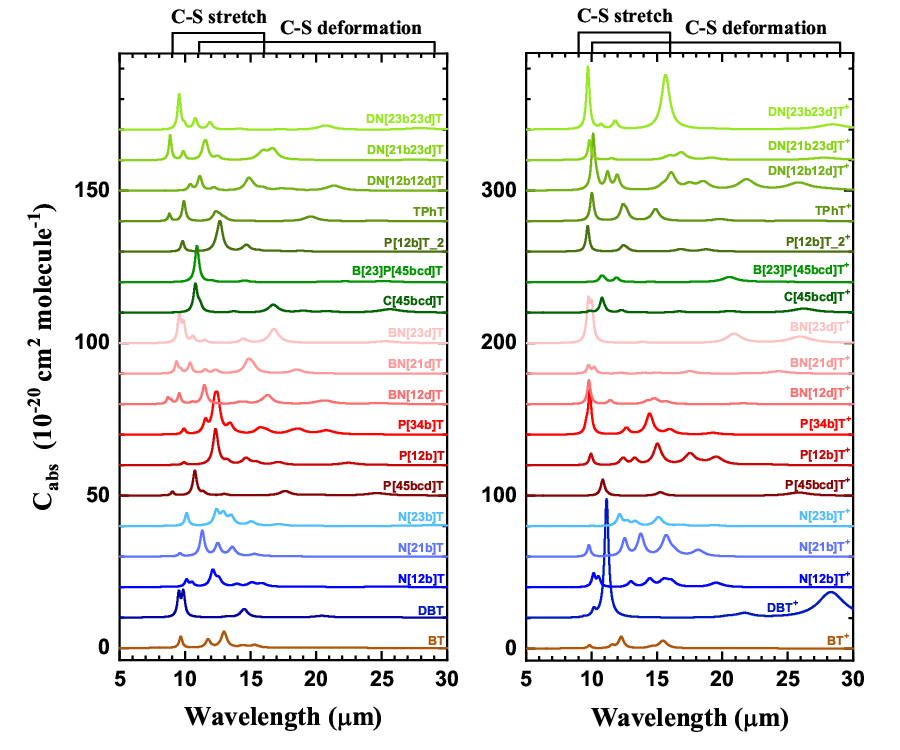,width=\textwidth}
  \end{center}
\vspace{-1mm}
\caption{\label{fig:CS_Total_spec}
              Calculated spectra of all the target molecules
              (left panel; see Figure~\ref{fig:TargetMolecules})
              and their cations (right panel).
              The wavelength ranges for the main vibrational
              bands are marked on top.
              }
\vspace{-0.1mm}
\end{figure*}

% Taking data from Tables~\ref{tab:CS_Freq_Int_Neutral}
% and \ref{tab:CS_Freq_Int_Cation},
To illustrate the C--S vibrations,
we show for each molecule
the C--S vibration only spectrum as follows.
We first convert the band strength ($A_{\tilde{\nu}}$)
given by the Gaussian outputs into absorption
cross section ($C_{\rm abs}$) through
\begin{equation}
\label{eq:epsilon2A}
\frac{A_{\tilde{\nu}}}{\km\mol^{-1}}
   = 100\times\int  \frac{\varepsilonWV}
              {\Liter\mol^{-1}\cm^{-1}}\,d\tilde{\nu}
   = 6.02\times10^{18}\times
       \int \frac{C_{\rm abs}(\tilde{\nu})}
            {\cm^2\mole^{-1}}\,d\tilde{\nu} ~~,
\end{equation}
where $\varepsilonWV$ is the molar extinction coefficient
and the integrations are performed over the band.
For each molecule, we approximate
each vibrational mode by a Drude function.
For the {\it j}-th mode,
let $\gamma_{j}$ be the FWHM ($\cm^{-1}$),
$\tilde{\nu}_{0,j}$ be the peak wavenumber ($\cm^{-1}$),
and $A_j$ be the calculated intensity ($\km\mol^{-1}$).
The total absorption cross section is then expressed as
\begin{equation}
\label{eq:Cabs_all}
C_{\rm abs}(\tilde{\nu})
= \sum_{j=1}^{N} \frac{1}{6.02\times10^{18}}\times A_{j,{\tilde{\nu}}}
\times \frac{2\gamma_j}{\pi}
\times \frac{1}{\left(\tilde{\nu}-\tilde{\nu}_{0,j}^{2}/
\tilde{\nu}\right)^2+\gamma_{j}^{2}} ~~,
\end{equation}
where $C_{\rm abs}$ in $\cm^2$ is the absorption cross section
given in eq.\,\ref{eq:epsilon2A} summed over N vibrational modes
of a molecule.
Using data from Tables~\ref{tab:CS_Freq_Int_Neutral}
and \ref{tab:CS_Freq_Int_Cation},
and taking $\gamma_{j} = 30\cm^{-1}$ for all modes,
we generate the absorption spectra ($C_{\rm abs}$)
for each molecule and its cation,
and plot in Figure~\ref{fig:CS_Total_spec}.\footnote{%
  Observationally, the 11.3$\mum$ C--H out-of-plane
  bending feature has a mean width of $\simali$0.36$\mum$
  (Draine \& Li 2007). This corresponds to $\simali$28$\cm^{-1}$.
}
When employing Drude functions to generate the spectra,
we take a wavenumber range of 0--3500$\cm^{-1}$,
with a step size of 0.2$\cm^{-1}$.
To better compare the generated spectra
with astronomical observations,
we plot in the x-axis the wavelength in unit of $\mum$
in the range of 5--30$\mum$,
where all the C--S vibrational bands occur.

A close inspection of Figure~\ref{fig:CS_Total_spec}
reveals that two cations, DBT+ and DN[23b23d]T+,
have very different spectra.
DBT+ has a very strong feature at 11.2$\mum$,
with an intensity of 220.5$\km\mol^{-1}$.
It arises from the asymmetric stretch of C--S--C
bonds in the five-membered ring.
BN[21d]T, BN[23d]T, DN[12b12d]T,
and DN[21b23d]T also have a similar configuration,
with a substructure just the same as DBT
but with more benzene rings attached.
They have similar vibrations, but occuring at shorter wavelength
and with much smaller intensities.
DBT+ also has a strong C--S deformation feature
at 28.3$\mum$, with an intensity of
47.65$\km\mol^{-1}$. This deformation,
the strongest C--S deformation vibration
among all of our target molecules,
originates from the twisting of C--S--C bonds
in the molecular plane.
DN[23b23d]T+ shows two strong bands
at 9.76 and 15.67$\mum$, both with intensities
exceeding 100$\km\mol^{-1}$.
The 9.76$\mum$ band also arises from
the asymmetric stretch of the C--S--C bonds,
while the 15.67$\mum$ feature arises from
the asymmetric stretch of the C--S--C
bonds coupled with the deformation of
the benzene ring.
In view of the fact that the spectra of these
two molecules are quite different from
that of other molecules, especially they
exhibit very strong vibrational intensities,
we will exclude these two molecules
(both neutral and cationic) in the following
so that our results will not be biased
by certain special molecules.

\subsection{Mean C--S Vibrational Spectra
                    \label{subsec:average_spectra}}
\begin{figure*}
 \hspace{-10mm}
  \begin{center}
  \epsfig{file=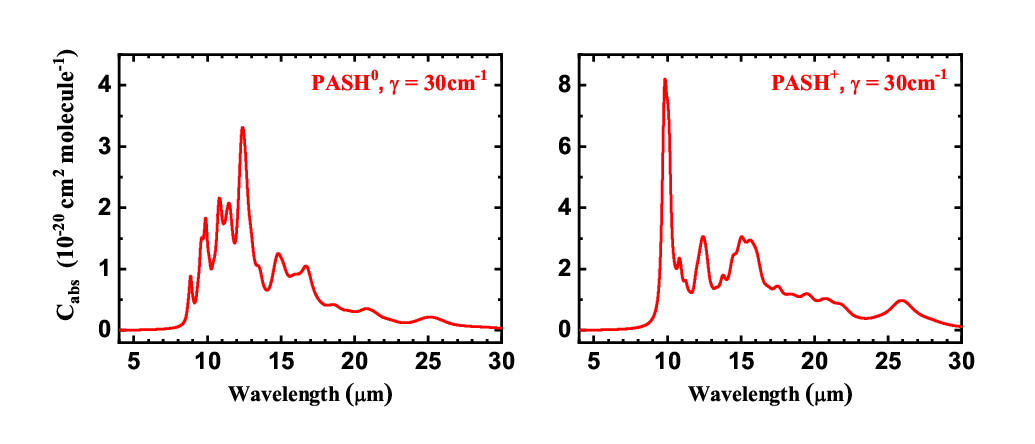,width=18cm}
  \end{center}
\vspace{-1mm}
\caption{\label{fig:average_spec_width30}
Average spectra of the C--S vibrations of all the target PASH
neutrals (left panel) and cations (right panel)
except DBT, DN[23b23d]T and their cations.
         }
\vspace{-0.1mm}
\end{figure*}

\begin{figure*}
 \hspace{-2mm}
  \begin{center}
  \epsfig{file=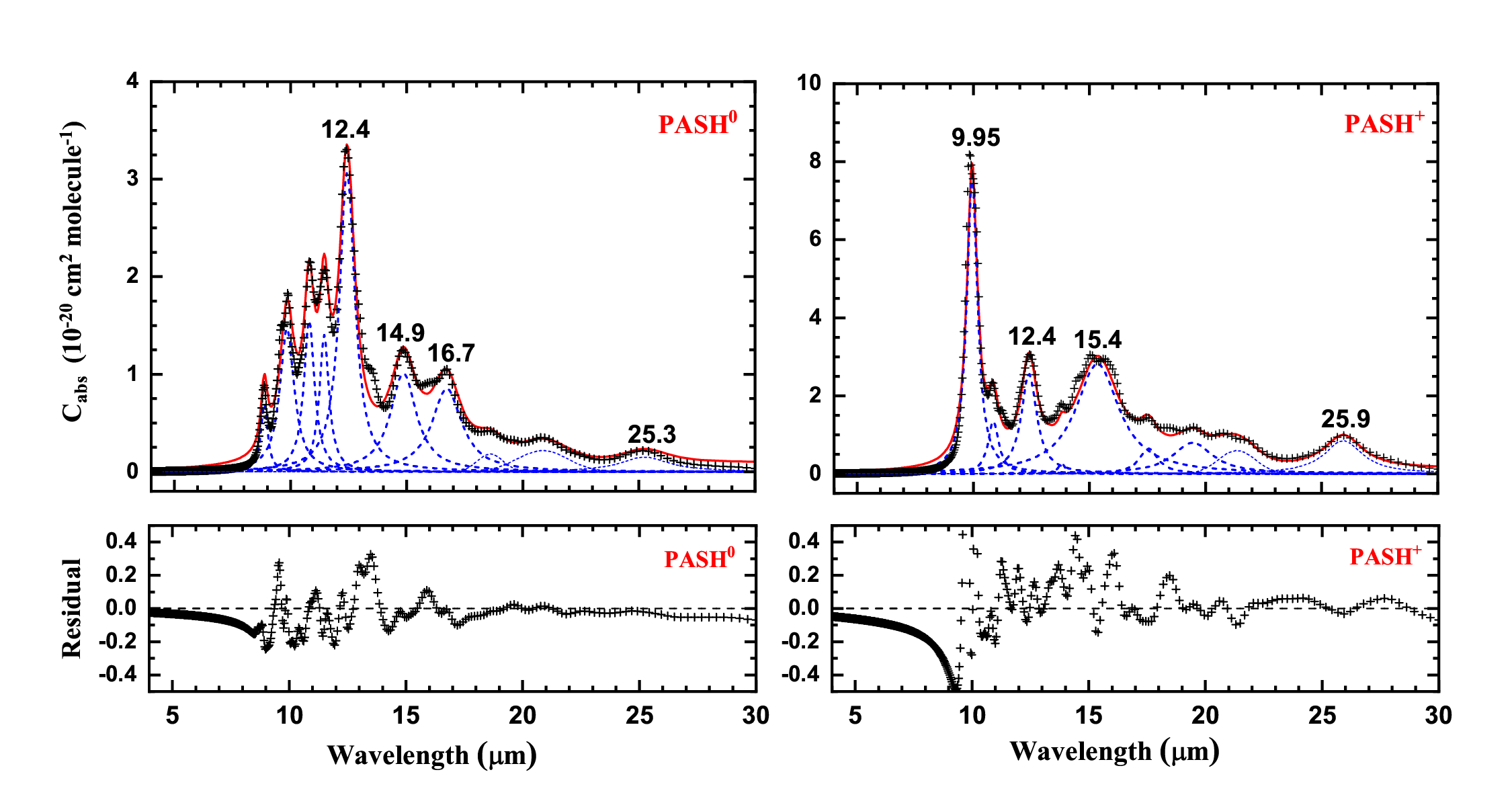,width=18cm}
  \end{center}
\vspace{-1mm}
\caption{\label{fig:FitResult}
  Fitting the average C--S vibrational spectra
  of neutrals (left panel) and cations (right panel)
  in terms of a linear background and several Drude
  and/or Gaussian functions.
  The central wavelengths (in $\mum$)
  of the most prominent bands are marked.
  Also shown are the fitting residuals
    obtained by subtracting the fitted spectra
    from the original spectra.
         }
\vspace{-0.1mm}
\end{figure*}

To highlight the C--S vibrational bands,
we calculate the mean spectra obtained
by averaging over all the target molecules
illustrated in Figure~\ref{fig:TargetMolecules}
(except DBT and DN[23b23d]T).
We show in Figure~\ref{fig:average_spec_width30}
the mean spectra for neutrals and cations,
and it is apparent that the average spectra exhibit
several bumps of different widths.
To facilitate future quantitative comparison
between these C--S vibrations with observations,
we fit each of the mean spectra (for neutrals or cations)
with a linear background plus several Lorentzian
and Gaussian functions.\footnote{%
  For the strong, sharp bands, we approximate
  them in terms of Lorentzian functions.
  In addition, there are a couple of broad
  ``plateau'' bands at $\simali$18.6 and
  20.9$\mum$ for neutrals and
  $\simali$21.4$\mum$ for cations
  arising from mixtures
  of neighboring vibrations.
  For these broad bands, we approximate them
  by Gaussian functions
  (see Tables~\ref{tab:FitResult_Neutral}, \ref{tab:FitResult_cation}).
  }
%
%Meanwhile, to release degree of freedom
%for the fitting, we take 5\% of the data
%points for the average spectra.
As illustrated in Figure~\ref{fig:FitResult},
a combination of Lorentz and Gaussian functions
closely reproduce the calculated mean spectra.
We tabulate in
Tables~\ref{tab:FitResult_Neutral} and \ref{tab:FitResult_cation}
the fitting results for the Lorentzian and Gaussian functions,
including peak wavelengths, FWHMs, and intensities
for neutrals and cations, respectively.

For neutral PASHs, we fit the mean spectrum with
eight Lorentz functions plus two Gaussian functions.
There are five relatively narrow bumps at wavelengths
$\lambda\simlt$\,14$\mum$,
with FWHMs all $\simlt$\,0.8$\mum$.
These bumps are all pure C--S stretches.
The other five bumps at wavelength of
$\lambda$\,$\simgt$\,14$\mum$
are mostly C--S deformation with
very minor contributions from C--S stretch.
The widths are relatively large, with FWHMs
all exceeding $\simgt$\,1.2$\mum$.
The large widths result from the broadening
of the mixtures of neighbouring vibrational frequencies,
suggesting a larger deviations in frequencies
for such vibrational modes.
Among all the C--S bands, the strongest is
the one at $\lambda$\,$\simali$\,12.44$\mum$,
with an integrated absorption cross section of
$\simali$$3.86\times10^{-25}\cm^3$ per S atom.
The two neighboring bands at $\lambda$\,$\simlt$\,14.86
and $\simali$16.73$\mum$ are prominent,
with an integrated absorption cross section
of $\simali$2.16 and $2.09\times10^{-25}\cm^3$
per S atom, respectively.

For cations, we consider eight Lorentz functions
and one Gaussian component for the fitting.
The four C--S stretch bumps at wavelength
$\simlt$\,14$\mum$ are also very narrow,
with FWHMs all below $\simlt0.8\mum$.
Among them, the bump at $\simlt$\,9.95$\mum$
is the strongest, with an integrated absorption
cross section of
$\simali$$6.60\times10^{-25}\cm^3$ per S atom.
The five C--S deformation bands at wavelengths
$\simgt$\,15$\mum$ are also very broad,
just as that for neutrals.
The FWHMs are mostly $\simali$2.0$\mum$,
except the very minor band at 17.55$\mum$.
Among them, the most prominent are the two
bands at $\simali$19.45$\mum$ and $\simlt$\,25.93$\mum$,
with an integrated absorption cross section of
$\simlt$\,2.58 and $2.63\times10^{-25}\cm^3$
per S atom, respectively.

\section{Astrophysical Implications}\label{sec:discussion}
\begin{figure*}
 \hspace{-2mm}
  \begin{center}
  \epsfig{file=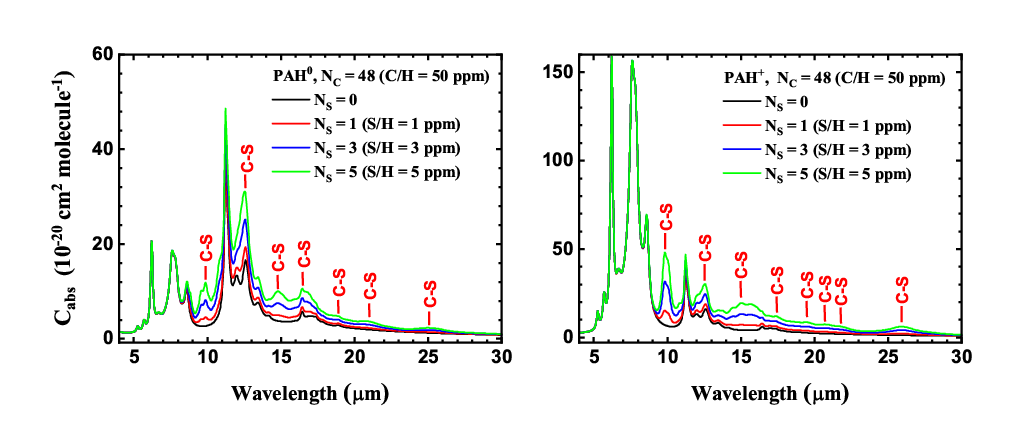,width=18cm}
  \end{center}
\vspace{-1mm}
\caption{\label{fig:PASH_PAH_C48H16}
  Absorption cross sections for a S-containing PAH
  molecule of 48 C atoms (left panel) and its cation
  (right panel). As described in eq.\,\ref{eq:Cabs_PAH_PASH},
  the absorption cross sections are
  obtained by adding the C--S vibrational bands
  of various S atoms ($N_{\rm S}$\,=\,0, 1, 3, and 5)
  to that of astro-PAHs of Draine \& Li (2007).
  Assuming the interstellar PAH population
  consumes a total C/H abundance of
  $\cpah=50\ppm$ and has a single size
  of $\NC=48$ C atoms, then, $\NS$ S atoms
  correspond to a total S/H depletion of
  $\spah=\cpah\times\left(\NS/\NC\right)
  \approx \NS\ppm$.
        }
\vspace{-0.1mm}
\end{figure*}

\begin{figure*}
 \hspace{-2mm}
  \begin{center}
  \epsfig{file=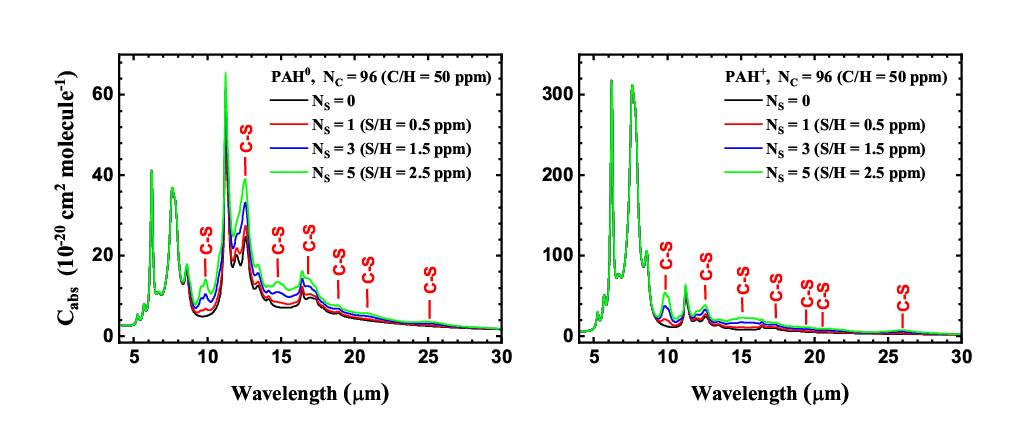,width=18cm}
  \end{center}
\vspace{-1mm}
\caption{\label{fig:PASH_PAH_C96H24}
    Same as Figure~\ref{fig:PASH_PAH_C48H16}
    but for a large molecule with $N_{\rm C}=96$.
         }
\vspace{-0.1mm}
\end{figure*}

Li \& Draine (2001) and Draine \& Li (2007) designed
an ``astro-PAH'' model by empirically synthesizing
$C_{\rm abs}({\rm PAH})$, a set of absorption cross
sections for PAHs that are consistent
with spectroscopic observations of PAH emission
in various astrophysical environments.
Although the resulting ``astro-PAH'' absorption
cross sections do not represent any specific material,
they are generally consistent with laboratory data
and approximate the actual absorption properties
of the PAH mixture in astrophysical regions.
For a sulfurated PAH molecule consisting of
$N_{\rm C}$ C atoms and $N_{\rm S}$ S atoms,
we approximate its absorption cross section
$C_{\rm abs}(N_{\rm C}, N_{\rm S})$
as a combination of the absorption cross section
of pure PAH of the same size
$C_{\rm abs}(N_{\rm C})$
and that of C--S vibrations:
\begin{equation}
    \label{eq:Cabs_PAH_PASH}
  C_{\rm abs}(N_{\rm C}, N_{\rm S})
  = C_{\rm abs}(N_{\rm C})
  +N_{\rm S}\times
  \left\{\frac{C_{\rm abs}({\rm PASH})}{N_{\rm S}}\right\} ~~.
 \end{equation}
where $C_{\rm abs}(N_{\rm C}, N_{\rm S})$ is the absorption
cross section of a sulfurated PAH molecule
of $N_{\rm C}$ C atoms and $N_{\rm S}$ S atoms,
$C_{\rm abs}(N_{\rm C})$ is the absorption
cross section of a pure PAH molecule
of $N_{\rm C}$ C atoms (see Li \& Draine 2001,
Draine \& Li 2007, Draine et al.\ 2021),
and $C_{\rm abs}({\rm PASH})/N_{\rm S}$
is the average cross section of C--S vibrations
(on a per S atom basis) calculated from
eq.\ref{eq:Cabs_all} for all the target molecules.
The resulting absorption cross sections
are shown in Figures~\ref{fig:PASH_PAH_C48H16}
and \ref{fig:PASH_PAH_C96H24}
for both neutrals and cations
with $N_{\rm C} = 48$ and $N_{\rm C} = 96$
and various S atoms, i.e.,  $N_{\rm S} = 0, 1, 3, 5$.

Figures~\ref{fig:PASH_PAH_C48H16}
and \ref{fig:PASH_PAH_C96H24}
clearly show that the C--S vibrational bands
at $\simali$10, 15 and 25$\mum$
are rather strong and not mixed with
the classical C--H and C--C bands of PAHs
at 3.3, 6.2, 7.7, 8.6, 11.3 and 12.7$\mum$.
It is apparent that with even one S atom
corporated into a PAH molecule,
the C--S band at $\simali$10$\mum$
is appreciable for both neutrals and cations.
Apparently, the $\simali$10$\mum$ C--S band
would be the most promising tracer for
interstellar PASH.
On the other hand, the C--S band at 15$\mum$
is broader and less significant.
It requires a larger degree of sulfuration,
e.g., $N_{\rm S}=3$,
for the 15$\mum$ C--S band
to be clearly seen in the vibrational spectra of neutrals.
This band is much weaker in cations and requires
$N_{\rm S}=5$ for the 15$\mum$ band
to be seen. The $25\mum$ bands in both neutrals
and cations are quite weak.

To account for the 3.3, 6.2, 7.7, 8.6, 11.3 and 12.7$\mum$
emission observed in the Galactic diffuse ISM,
Li \& Draine (2001) and Draine \& Li (2007)
found that an amount of C/H\,$\approx$40--60$\ppm$
is required to be locked up in PAHs.
%{\bf
Let's take an intermediate value of
$\cpah=50\ppm$ and assume a single
PAH size of $\NC$ C atoms for the entire
interstellar PAH population which depletes
a total C/H abundance of $\cpah$.
Then, such a PAH population containing
$\NS$ S atoms would lock up a total S/H
abundance of
$\spah=\cpah\times\left(\NS/\NC\right)$.
As shown in Figure~\ref{fig:PASH_PAH_C48H16},
PAH molecules of $\NC=48$ C atoms
and $\NS=1$ S atom exhibit a noticeable
C--S band at 10$\mum$.
For the Galactic ISM,  Li \& Draine (2001)
derived a mean PAH size of $\NC\approx 100$.
In this case,  $\NS=1$ corresponds to
$\spah\approx0.5\ppm$.
%}

%{\bf
We note that, while the IR spectra
of several regions in the Orion Bar
recently obtained by the {\it Mid-Infrared Instrument}
(MIRI) on board JWST (Chown et al.\ 2024)
do not seem to show any emission at 10$\mum$,
the {\it Spitzer Infrared Spectrograph} (IRS) spectra
of the Galactic reflection nebulae
NGC~2023 and NGC~7023 do exhibit a weak
band at 10.1$\mum$ (see Werner et al.\ 2004,
Sellgren et al.\ 2010).\footnote{%
%  {\bf
    Werner et al.\ (2004) attributed the weak
    10.1$\mum$ emission band seen in NGC~7023
    to the C--H oop bending modes of PAHs
    (Hony et al.\ 2001).
  }
%  }
We should also note that the 10$\mum$ bands
shown in Figures~\ref{fig:PASH_PAH_C48H16}
and \ref{fig:PASH_PAH_C96H24} are the absorption
cross section of the C--S stretch and should not
be {\it readily} used to estimate the amount of S/H
depleted in PAHs by {\it directly} comparing with
the astronomical spectra.
To quantitatively explore the amount of S/H
locked up in PASHs, we have to fold the absorption
cross sections of PASHs with the starlight radiation
field to model the vibrational excitation of PASHs
and calculate their IR emission spectra.
This will be investigated in a subsequent paper.
%}

%{\bf
Finally, we note that the detectability of the 10$\mum$
C--S band could be affected by the broad silicate
absorption band centered at 9.7$\mum$.
In external galaxies such as M82, NGC~253 and Circinus,
their {\it Spitzer}/IRS spectra show a strong 9.7$\mum$
silicate absorption band (see Sturm et al.\ 2000).
The 10$\mum$ C--S band of PASHs could be hidden
by the silicate absorption band even if PASHs are present.
This could also be true in the Orion Bar
which subjects to a substantial amount of extinction
(see Fuente et al.\ 2024).
In the subsequent paper we will also investigate this
in detail, in combination with the latest available JWST
spectroscopic data.
%}

\section{Summary}\label{sec:summary}
Despite its biogeneic and astrochemical importance,
S-containing molecules and ices in molecular clouds
are far below expected from the interstellar S abundance
 of $\sism\approx22.4\ppm$.
In the diffuse ISM, although it was previously thought
that S is not depleted, recent observations suggest
that an appreciable portion of the interstellar S
in the diffuse ISM is also missing from the gas.
With common sulfides like MgS, FeS and SiS$_2$
excluded as a major S reservior, we have examined
the possibility that the missing S may have partly
been locked up in PASHs---polycyclic aromatic sulfur
heterocycles by computing the IR vibrational spectra
of 18 representative PASH species.
We find that these molecules exhibit a prominent,
stand-alone C--S stretching band at $\simali$10$\mum$
which is detectable by JWST if several ppm of S/H
is depleted in PASHs. A weak emission band
at 10.1$\mum$ has already been seen
in the {\it Spitzer}/IRS spectra of NGC~2023
and NGC~7023. Future JWST observations will
allow one to quantitatively examine
if PASH could account for some portions
of the S elements missing from the gas phase.

\acknowledgments{%
We dedicate this paper to the memory
of Edward B. Jenkins with whom we
discussed the possible depletion routes
of S during the BruceFest in Florence, Italy
on October 30 -- November 3, 2023.
Ed was an active participant
in the biweekly Interstellar Medium Seminar
one of us (AL) once organized and benefited
much in Princeton in 1999--2002.
We thank the anonymous referees
for helpful comments and suggestions.
We thank B.M.~Broderick, B.T.~Draine,
and E.F.~van Dishoeck for stimulating discussions.
XJY and HLJ are supported in part by
NSFC~12333005 and 12122302
and CMS-CSST-2021-A09.
%AL is supported in part by NASA NNX13AE63G;
}

%%%%%%%%%%%%%%% References %%%%%%%%%%%%%%%%%%%%%%%%%%%%

\clearpage

%%%%%%%%%% Figure 1 %%%%%%%%%%%
%%%%%%%%%% Figure 1 %%%%%%%%%%%

%%%%%%%%%% Figure 2 %%%%%%%%%%%
%%%%%%%%%% Figure 2 %%%%%%%%%%%

%%%%%%%%%% Figure 3 %%%%%%%%%%%
%%%%%%%%%% Figure 3 %%%%%%%%%%%

%%%%%%%%%% Figure 4 %%%%%%%%%%%
%%%%%%%%%% Figure 4 %%%%%%%%%%%

%%%%%%%%%% Figure 5 %%%%%%%%%%%
%%%%%%%%%% Figure 5 %%%%%%%%%%%

%%%%%%%%%% Figure 6 %%%%%%%%%%%
%%%%%%%%%% Figure 6 %%%%%%%%%%%

%%%%%%%%%% Figure 7 %%%%%%%%%%%
%%%%%%%%%% Figure 7 %%%%%%%%%%%

%%%%%%%%%% Figure 8 %%%%%%%%%%%
%%%%%%%%%% Figure 8 %%%%%%%%%%%

%%%%%%%%%% Figure 9 %%%%%%%%%%%
%%%%%%%%%% Figure 9 %%%%%%%%%%%

\clearpage

%========Table 1=============
\begin{table}
  {\scriptsize
  \centering
  \caption{Interstellar Sulfur Abundance}
  \label{tab:Sabundance}
  \begin{tabular}[c]{cccccccccc}
\noalign{\smallskip} \noalign{\smallskip}
  \toprule[1pt]%

Elemental & B and F, G Stars$^a$ &  Sun$^b$  &  Proto-Sun$^c$ & Proto-Sun  &
\multicolumn{2}{c}{Diffuse ISM (ppm)} & ~ &
                                            \multicolumn{2}{c}{Molecular
                                            Clouds (ppm)}   \\
    \cline{6-7}\cline{9-10}\noalign{\smallskip}
 Abundance & (ppm) & (ppm) & (ppm) & +GCE$^d$ (ppm) & Gas$^e$ & Dust$^f$ &
                                                                       ~
           & Gas$^g$ & Dust$^h$ \\
\noalign{\smallskip}  \hline \noalign{\smallskip}
S/H  & 14.5$\pm$1.0 & 16.2$\pm$1.3  &  18.2$\pm$1.7  &  22.4$\pm$1.7  &
 9.6  & 12.8 & ~ & $<$ 1  & 22.4  \\
  \bottomrule[1pt]\noalign{\smallskip}\noalign{\smallskip}%
  \end{tabular}
  }
  \scriptsize{$a$: Asplund et al.\ 2009}\\
  \scriptsize{$b$: Przylilla et al.\ 2008}\\
  \scriptsize{$c$: Lodders 2003}\\
  \scriptsize{$d$: Chiappini et al.\ 2003}\\
  \scriptsize{$e$: Draine \& Hensley 2021}\\
  \scriptsize{$f$: Our adopted sulfur abundance of 22.4\,ppm
                            minus that in the gas phase}\\
  \scriptsize{$g$: Fuente et al. 2019}\\
  \scriptsize{$h$: Our adopted sulfur abundance of 22.4\,ppm
                            minus that in the gas phase}\\
 \end{table}
% =================Table 1 ==================

%=================Table 2 ==================
\onecolumn
\begin{longtable}[c]{cccccc}
  \caption[]{
  Frequencies and Intensities of the C--S Vibrational Modes
  of Neutral PASHs}
\label{tab:CS_Freq_Int_Neutral}\\
\noalign{\smallskip} \noalign{\smallskip}
		\hline \hline
		\multirow{3}{*}{Molecule} & \multicolumn{2}{c}{C--S Stretch} &~& \multicolumn{2}{c}{C--S Deformation}  \\ \cline{2-3}\cline{5-6}
		{~~~~~~~~~~~~~~~~~~~~~} & \makecell[c]{Wavelength \\ ($\mu$m)} & \makecell[c]{Intensity \\{(km\,mol$^{-1}$)}} &~  & \makecell[c]{Wavelength \\ ($\mu$m)} & \makecell[c]{Intensity \\(km\,mol$^{-1}$)} \\
		\hline
		\endfirsthead

\hline
		\multirow{3}{*}{Molecule} & \multicolumn{2}{c}{C--S Stretch} &~& \multicolumn{2}{c}{C--S Deformation}  \\ \cline{2-3}\cline{5-6}
		{~~~~~~~~~~~~~~~~~~~~~} & \makecell[c]{Wavelength \\ ($\mu$m)} & \makecell[c]{Intensity \\{(km\,mol$^{-1}$)}} &~  & \makecell[c]{Wavelength \\ ($\mu$m)} & \makecell[c]{Intensity \\(km\,mol$^{-1}$)} \\
		\hline
		\endhead

		\hline
		\endfoot

		\hline\hline
		\endlastfoot
		\multirow{4}{*}{BT}	&	9.67 	&	10.42 	&	~	&	15.33 	&	2.46 	\\
		~	&	11.76 	&	7.52 	&	~	&	19.37 	&	0.33 	\\
		~	&	12.97 	&	14.73 	&	~	&	20.94 	&	0.39 	\\
		~	&	14.43 	&	1.97 	&	~	&		&		\\
		\midrule[0.1pt]%
		\multirow{6}{*}{DBT}	&	9.51 	&	11.33 	&	~	&	20.42 	&	1.28 	\\
		~	&	9.54 	&	10.55 	&	~	&	20.74 	&	0.23 	\\
		~	&	9.87 	&	22.54 	&	~	&	24.50 	&	0.24 	\\
		~	&	13.21 	&	0.65 	&	~	&		&		\\
		~	&	14.44 	&	1.20 	&	~	&		&		\\
		~	&	14.52 	&	6.55 	&	~	&		&		\\
		\midrule[0.1pt]%
		\multirow{5}{*}{N[12b]T}	&	10.12 	&	6.41 	&	~	&	15.10 	&	3.54 	\\
		~	&	10.54 	&	3.71 	&	~	&	15.87 	&	2.90 	\\
		~	&	12.12 	&	14.65 	&	~	&	18.29 	&	0.34 	\\
		~	&	12.57 	&	6.80 	&	~	&	25.48 	&	0.25 	\\
		~	&	13.96 	&	2.17 	&	~	&		&		\\
		\midrule[0.1pt]%
		\multirow{4}{*}{N[21b]T}	&	9.62 	&	2.93 	&	~	&	15.31 	&	1.99 	\\
		~	&	11.31 	&	23.65 	&	~	&	16.15 	&	0.38 	\\
		~	&	12.50 	&	11.47 	&	~	&	18.09 	&	0.04 	\\
		~	&	13.61 	&	8.17 	&	~	&		&		\\
		\midrule[0.1pt]%
		\multirow{5}{*}{N[23b]T}	&	10.12 	&	11.94 	&	~	&	17.15 	&	1.61 	\\
		~	&	12.40 	&	13.41 	&	~	&	18.92 	&	0.21 	\\
		~	&	12.93 	&	10.21 	&	~	&	28.96 	&	0.85 	\\
		~	&	13.52 	&	8.14 	&	~	&		&		\\
		~	&	15.07 	&	3.71 	&	~	&		&		\\
		\midrule[0.1pt]%
		\multirow{7}{*}{P[45bcd]T}	&	9.04 	&	3.54 	&	~	&	13.01 	&	1.30 	\\
		~	&	10.75 	&	22.75 	&	~	&	15.18 	&	0.39 	\\
		~	&	11.40 	&	2.66 	&	~	&	17.64 	&	4.07 	\\
		~	&	14.03 	&	0.19 	&	~	&	19.80 	&	0.23 	\\
		~	&		&		&	~	&	20.18 	&	0.60 	\\
		~	&		&		&	~	&	24.52 	&	0.96 	\\
		~	&		&		&	~	&	24.77 	&	1.25 	\\
		\midrule[0.1pt]%
		\multirow{5}{*}{P[12b]T}	&	9.92 	&	2.42 	&	~	&	15.40 	&	2.37 	\\
		~	&	12.33 	&	32.40 	&	~	&	17.13 	&	2.57 	\\
		~	&	12.37 	&	0.89 	&	~	&	17.91 	&	0.03 	\\
		~	&	13.22 	&	3.19 	&	~	&	22.48 	&	2.01 	\\
		~	&	14.66 	&	5.92 	&	~	&		&		\\
		\midrule[0.1pt]%
		\multirow{11}{*}{P[34b]T}	&	9.92 	&	5.01 	&	~	&	11.56 	&	11.33 	\\
		~	&	13.44 	&	8.15 	&	~	&	12.26 	&	23.68 	\\
		~	&		&		&	~	&	12.50 	&	25.49 	\\
		~	&		&		&	~	&	14.16 	&	0.25 	\\
		~	&		&		&	~	&	15.70 	&	4.78 	\\
		~	&		&		&	~	&	16.06 	&	1.16 	\\
		~	&		&		&	~	&	16.24 	&	2.03 	\\
		~	&		&		&	~	&	18.14 	&	1.71 	\\
		~	&		&		&	~	&	18.67 	&	4.17 	\\
		~	&		&		&	~	&	19.54 	&	0.17 	\\
		~	&		&		&	~	&	20.81 	&	3.51 	\\
		\midrule[0.1pt]%
		\multirow{7}{*}{BN[12d]T}	&	8.69 	&	5.50 	&	~	&	14.54 	&	1.52 	\\
		~	&	8.92 	&	3.45 	&	~	&	15.77 	&	0.58 	\\
		~	&	9.56 	&	9.79 	&	~	&	16.31 	&	8.10 	\\
		~	&	10.57 	&	1.53 	&	~	&	20.23 	&	1.12 	\\
		~	&	11.48 	&	17.51 	&	~	&	20.77 	&	2.57 	\\
		~	&	12.32 	&	0.14 	&	~	&	24.84 	&	0.86 	\\
		~	&	14.12 	&	1.32 	&	~	&		&		\\
		\midrule[0.1pt]%
		\multirow{7}{*}{BN[21d]T}	&	9.35 	&	10.82 	&	~	&	11.54 	&	3.36 	\\
		~	&	9.66 	&	3.18 	&	~	&	17.39 	&	0.16 	\\
		~	&	10.39 	&	9.68 	&	~	&	18.55 	&	3.53 	\\
		~	&	12.33 	&	3.39 	&	~	&	20.82 	&	0.05 	\\
		~	&	13.95 	&	0.12 	&	~	&	24.18 	&	0.02 	\\
		~	&	14.80 	&	10.40 	&	~	&		&		\\
		~	&	15.15 	&	6.37 	&	~	&		&		\\
		\midrule[0.1pt]%\\\\%
		\multirow{2}{*}{BN[23d]T}	&	9.54 	&	18.27 	&	~	&	11.50 	&	2.57 	\\
		~	&	9.62 	&	7.58 	&	~	&	14.47 	&	3.86 	\\
		\multirow{3}{*}{BN[23d]T}	&	9.88 	&	17.15 	&	~	&	16.79 	&	13.02 	\\
		~	&	10.60 	&	5.64 	&	~	&	20.78 	&	0.34 	\\
		~	&		&		&	~	&	25.28 	&	1.70 	\\
		\midrule[0.1pt]%
		\multirow{7}{*}{C[45bcd]T}	&	10.79 	&	25.80 	&	~	&	13.74 	&	1.12 	\\
		~	&	11.14 	&	6.29 	&	~	&	16.71 	&	5.19 	\\
		~	&		&		&	~	&	16.82 	&	2.17 	\\
		~	&		&		&	~	&	19.03 	&	1.46 	\\
		~	&		&		&	~	&	20.59 	&	0.53 	\\
		~	&		&		&	~	&	21.09 	&	0.63 	\\
		~	&		&		&	~	&	25.64 	&	3.10 	\\
		\midrule[0.1pt]%
		\multirow{4}{*}{B[23]P[45bcd]T}	&	10.88 	&	14.17 	&	~	&	20.51 	&	0.28 	\\
		~	&	10.93 	&	20.04 	&	~	&	22.21 	&	0.71 	\\
		~	&	12.04 	&	1.08 	&	~	&	25.15 	&	0.94 	\\
		~	&	14.56 	&	1.57 	&	~	&		&		\\
		\midrule[0.1pt]%
		\multirow{5}{*}{P[12b]T\_2}	&	9.82 	&	9.71 	&	~	&	16.98 	&	0.68 	\\
		~	&	12.40 	&	4.15 	&	~	&	18.83 	&	0.77 	\\
		~	&	12.67 	&	26.67 	&	~	&		&		\\
		~	&	14.69 	&	5.92 	&	~	&		&		\\
		~	&	15.58 	&	0.95 	&	~	&		&		\\
		\midrule[0.1pt]%
		\multirow{4}{*}{TPhT}	&	8.81 	&	6.67 	&	~	&	12.64 	&	3.97 	\\
		~	&	9.91 	&	18.76 	&	~	&	13.00 	&	2.38 	\\
		~	&	12.33 	&	7.47 	&	~	&	19.60 	&	4.42 	\\
		~	&	14.77 	&	0.53 	&	~	&	24.73 	&	0.47 	\\
		\midrule[0.1pt]%
		\multirow{5}{*}{DN[12b12d]T}	&	10.41 	&	5.50 	&	~	&	14.85 	&	9.90 	\\
		~	&	11.13 	&	13.32 	&	~	&	17.34 	&	1.46 	\\
		~	&	12.21 	&	2.74 	&	~	&	18.09 	&	1.19 	\\
		~	&	15.10 	&	2.68 	&	~	&	21.37 	&	4.63 	\\
		~	&	15.91 	&	2.31 	&	~	&		&		\\
		\midrule[0.1pt]%
		\multirow{5}{*}{DN[21b23d]T}	&	8.86 	&	23.09 	&	~	&	15.65 	&	1.27 	\\
		~	&	9.87 	&	8.06 	&	~	&	15.97 	&	6.55 	\\
		~	&	11.43 	&	8.79 	&	~	&	16.70 	&	9.71 	\\
		~	&	11.61 	&	13.01 	&	~	&	18.98 	&	0.23 	\\
		~	&	12.49 	&	3.15 	&	~	&	27.57 	&	0.60 	\\
		\midrule[0.1pt]%
		\multirow{5}{*}{DN[23b23d]T}	&	9.55 	&	32.73 	&	~	&	14.15 	&	1.12 	\\
		~	&	9.95 	&	3.93 	&	~	&	17.24 	&	0.44 	\\
		~	&	10.78 	&	10.07 	&	~	&	19.44 	&	0.01 	\\
		~	&	11.90 	&	6.80 	&	~	&	20.75 	&	4.04 	\\
		~	&		&		&	~	&	27.94 	&	1.58 	\\
%\noalign{\smallskip}
\\
\end{longtable}

%===========Table 2==============

%===========Table 3==============
\onecolumn
\begin{longtable}[c]{cccccc}
  \caption[]{
  Same as Table~\ref{tab:CS_Freq_Int_Neutral} but for Cationic PASHs}
	\label{tab:CS_Freq_Int_Cation}\\
	
	\hline \hline
	\multirow{3}{*}{Molecule} & \multicolumn{2}{c}{C-S Stretch} &~& \multicolumn{2}{c}{C-S Deformation}  \\ \cline{2-3}\cline{5-6}
	{~~~~~~~~~~~~~~~~~~~~~~} & \makecell[c]{Wavelength \\ ($\mu$m)} & \makecell[c]{Intensity \\(km\,mol$^{-1}$)} &~  & \makecell[c]{Wavelength \\ ($\mu$m)} & \makecell[c]{Intensity \\(km\,mol$^{-1}$)} \\
	\hline
	\endfirsthead

	\hline
	\multirow{3}{*}{Molecular} & \multicolumn{2}{c}{C-S Stretch} &~& \multicolumn{2}{c}{C-S Deformation}  \\ \cline{2-3}\cline{5-6}
	{~~~~~~~~~~~~~~~~~~~~~~} & \makecell[c]{Wavelength \\ ($\mu$m)} & \makecell[c]{Intensity \\(km\,mol$^{-1}$)} &~  & \makecell[c]{Wavelength \\ ($\mu$m)} & \makecell[c]{Intensity \\(km\,mol$^{-1}$)} \\
	\hline
	\endhead

	\hline
	\endfoot

	\hline\hline
	\endlastfoot

	\multirow{4}{*}{$\rm{BT}^+$}	&	9.84 	&	5.83 	&	~	&	15.47 	&	13.67 	\\
		~	&	11.59 	&	5.41 	&	~	&	21.90 	&	0.19 	\\
		~	&	12.28 	&	20.84 	&	~	&		&		\\
		~	&	14.58 	&	2.49 	&	~	&		&		\\
		\midrule[0.1pt]%
		\multirow{4}{*}{$\rm{DBT}^+$}	&	9.49 	&	1.64 	&	~	&	20.48 	&	2.18 	\\
		~	&	10.19 	&	12.46 	&	~	&	21.72 	&	7.63 	\\
		~	&	11.15 	&	220.54 	&	~	&	28.32 	&	47.63 	\\
		~	&	14.26 	&	0.49 	&	~	&		&		\\
		\midrule[0.1pt]%
		\multirow{6}{*}{$\rm{N[12b]T}^+$}	&	14.46 	&	14.76 	&	~	&	10.16 	&	23.79 	\\
		~	&		&		&	~	&	10.55 	&	16.87 	\\
		~	&		&		&	~	&	13.02 	&	10.11 	\\
		~	&		&		&	~	&	15.55 	&	11.87 	\\
		~	&		&		&	~	&	16.09 	&	10.42 	\\
		~	&		&		&	~	&	19.54 	&	8.16 	\\
		\midrule[0.1pt]%
		\multirow{4}{*}{$\rm{N[21b]T}^+$}	&	9.80 	&	21.31 	&	~	&	15.71 	&	36.99 	\\
		~	&	11.33 	&	0.81 	&	~	&	16.32 	&	6.59 	\\
		~	&	12.54 	&	32.11 	&	~	&	18.18 	&	11.53 	\\
		~	&	13.77 	&	40.46 	&	~	&		&		\\
		\midrule[0.1pt]%
		\multirow{5}{*}{$\rm{N[23b]T}^+$}	&	10.04 	&	0.95 	&	~	&	16.60 	&	0.80 	\\
		~	&	12.15 	&	21.09 	&	~	&	17.02 	&	1.63 	\\
		~	&	12.73 	&	7.22 	&	~	&	19.40 	&	2.02 	\\
		~	&	13.36 	&	7.89 	&	~	&		&		\\
		~	&	15.13 	&	16.84 	&	~	&		&		\\
		\midrule[0.1pt]%
		\multirow{3}{*}{$\rm{P[45bcd]T}^+$}	&	10.86 	&	29.59 	&	~	&	20.19 	&	0.37 	\\
		~	&	11.23 	&	1.08 	&	~	&	25.84 	&	5.92 	\\
		~	&	15.28 	&	6.38 	&	~	&		&		\\
		\midrule[0.1pt]%
		\multirow{5}{*}{$\rm{P[12b]T}^+$}	&	9.97 	&	21.01 	&	~	&	15.49 	&	1.61 	\\
		~	&	12.28 	&	3.17 	&	~	&	17.54 	&	20.33 	\\
		~	&	12.45 	&	10.75 	&	~	&	18.45 	&	2.36 	\\
		~	&	13.32 	&	11.57 	&	~	&	19.55 	&	13.95 	\\
		~	&	15.05 	&	38.09 	&	~	&	23.16 	&	1.71 	\\
		\midrule[0.1pt]%
		\multirow{4}{*}{$\rm{P[34b]T}^+$}	&	9.61 	&	13.37 	&	~	&	15.99 	&	5.66 	\\
		~	&	9.87 	&	79.21 	&	~	&	16.03 	&	3.18 	\\
		~	&	12.67 	&	12.13 	&	~	&	19.29 	&	3.15 	\\
		~	&	14.45 	&	38.65 	&	~	&		&		\\
		\midrule[0.1pt]%
		\multirow{6}{*}{$\rm{BN[12d]T}^+$}	&	9.80 	&	43.65 	&	~	&	21.63 	&	1.58 	\\
		~	&	9.93 	&	0.86 	&	~	&		&		\\
		~	&	11.45 	&	7.47 	&	~	&		&		\\
		~	&	14.26 	&	4.59 	&	~	&		&		\\
		~	&	14.83 	&	9.35 	&	~	&		&		\\
		~	&	15.72 	&	4.09 	&	~	&		&		\\
		\midrule[0.1pt]%
		\multirow{6}{*}{$\rm{BN[21d]T}^+$}	&	9.69 	&	0.40 	&	~	&	17.52 	&	4.62 	\\
		~	&	9.78 	&	14.73 	&	~	&	21.43 	&	0.14 	\\
		~	&	10.22 	&	11.04 	&	~	&	24.33 	&	4.33 	\\
		~	&	12.24 	&	1.67 	&	~	&		&		\\
		~	&	14.03 	&	1.47 	&	~	&		&		\\
		~	&	15.01 	&	2.02 	&	~	&		&		\\
		\midrule[0.1pt]%
		\multirow{6}{*}{$\rm{BN[23d]T}^+$}	&	9.56 	&	4.83 	&	~	&	15.92 	&	0.42 	\\
		~	&	9.76 	&	68.31 	&	~	&	20.91 	&	17.14 	\\
		~	&	10.02 	&	63.42 	&	~	&	25.94 	&	11.85 	\\
		~	&	10.57 	&	0.56 	&	~	&		&		\\
		~	&	11.49 	&	0.47 	&	~	&		&		\\
		~	&	13.92 	&	2.48 	&	~	&		&		\\
		\midrule[0.1pt]%
		\multirow{5}{*}{$\rm{C[45bcd]T}^+$}	&	9.88 	&	2.94 	&	~	&	16.74 	&	2.28 	\\
		~	&	10.81 	&	26.04 	&	~	&	20.58 	&	2.76 	\\
		~	&	11.01 	&	3.11 	&	~	&	26.25 	&	7.26 	\\
		~	&	12.30 	&	4.36 	&	~	&		&		\\
		~	&	13.64 	&	0.95 	&	~	&		&		\\
		\midrule[0.1pt]%
		\multirow{4}{*}{$\rm{B[23]P[45bcd]T}^+$}	&	10.76 	&	9.11 	&	~	&	20.56 	&	9.33 	\\
		~	&	10.95 	&	5.97 	&	~	&	22.14 	&	0.54 	\\
		~	&	11.93 	&	8.18 	&	~	&	25.68 	&	0.17 	\\
		~	&	14.55 	&	0.39 	&	~	&		&		\\
		\midrule[0.1pt]%
		\multirow{1}{*}{$\rm{P[12b]T\_2}^+$}	&	9.71 	&	47.36 	&	~	&	15.52 	&	0.14 	\\
		\multirow{3}{*}{$\rm{P[12b]T\_2}^+$}	&	12.42 	&	9.76 	&	~	&	16.86 	&	4.42 	\\
		~	&	12.70 	&	4.22 	&	~	&	18.77 	&	3.90 	\\
		~	&	14.95 	&	0.43 	&	~	&		&		\\
		\midrule[0.1pt]%
		\multirow{5}{*}{$\rm{TPhT}^+$}	&	10.02 	&	40.85 	&	~	&	19.81 	&	3.65 	\\
		~	&	10.09 	&	12.65 	&	~	&	26.30 	&	0.72 	\\
		~	&	12.40 	&	24.61 	&	~	&		&		\\
		~	&	12.67 	&	13.84 	&	~	&		&		\\
		~	&	14.90 	&	21.82 	&	~	&		&		\\
		\midrule[0.1pt]%
		\multirow{6}{*}{$\rm{DN[12b12d]}T^+$}	&	10.13 	&	100.66 	&	~	&	17.50 	&	9.59 	\\
		~	&	10.40 	&	13.49 	&	~	&	18.56 	&	14.14 	\\
		~	&	11.23 	&	30.79 	&	~	&	21.86 	&	19.39 	\\
		~	&	11.97 	&	26.20 	&	~	&	25.84 	&	13.73 	\\
		~	&	15.45 	&	5.64 	&	~	&	27.98 	&	1.63 	\\
		~	&	16.10 	&	30.43 	&	~	&		&		\\
		\midrule[0.1pt]%
		\multirow{4}{*}{$\rm{DN[21b23d]T}^+$}	&	9.85 	&	36.79 	&	~	&	15.66 	&	1.42 	\\
		~	&	10.75 	&	0.39 	&	~	&	16.87 	&	13.15 	\\
		~	&	11.55 	&	2.88 	&	~	&	19.19 	&	3.91 	\\
		~	&	15.98 	&	5.36 	&	~	&	27.81 	&	4.13 	\\
		\midrule[0.1pt]%
		\multirow{4}{*}{$\rm{DN[23b23d]T}^+$}	&	9.73 	&	116.28 	&	~	&	17.06 	&	0.40 	\\
		~	&	10.75 	&	7.06 	&	~	&	19.29 	&	0.63 	\\
		~	&	11.82 	&	15.24 	&	~	&	28.43 	&	9.32 	\\
		~	&	15.67 	&	101.15 	&	~	&		&		\\
\end{longtable}

%========Table 3==============

%========Table 4 ==============
	\begin{table}
		\centering
		\caption{Parameters for Fitting the C--S Bands of Neutral PASHs. The band intensities
(cm$^3$ per S atom) is the absorption cross section of the band integrated over wavelength (i.e.,
$\int_{\rm band} C_{\rm abs}({\lambda}) d{\lambda}$)}

		\label{tab:FitResult_Neutral}
		\begin{tabular}[c]{ccc}
\noalign{\smallskip} \noalign{\smallskip}
		\toprule[1pt]%
		\makecell[c]{Wavelength \\ ($\mu$m)} & \makecell[c]{FWHM \\ ($\mu$m)} & \makecell[c]{Intensity \\($10^{-25}\cm^3$ per S)}  \\
		\hline
8.87 	&	0.35	&	0.38 	\\
9.85 	&	0.75	&	1.73 	\\
10.80 	&	0.56	&	1.36 	\\
11.45 	&	0.50  	&	1.10 	\\
12.44 	&	0.80    &	3.86 	\\
14.86 	&	1.36	&	2.16 	\\
16.73 	&	1.56	&	2.09 	\\
18.60* 	&	1.29	&	0.25 	\\
20.85* 	&	2.56	&	0.59 	\\
25.25 	&	2.56	&	0.60 	\\
		\bottomrule[1pt]%
		{*Gaussian Component}
		\end{tabular}
	\end{table}
% ========Table 4 ==============

% ========Table 5 ==============
	\begin{table}
		\centering
		\caption{Same as Table~\ref{tab:FitResult_Neutral} but
                  for PASH Cations}
		\label{tab:FitResult_cation}
		\renewcommand\arraystretch{1.2}%
		\setlength{\tabcolsep}{2.5mm}
		\begin{tabular}[c]{ccc}
\noalign{\smallskip} \noalign{\smallskip}
		\toprule[1pt]%
		\makecell[c]{Wavelength \\ ($\mum$)} & \makecell[c]{FWHM \\ ($\mu$m)}  & \makecell[c]{Intensity \\($10^{-25}\cm^3$ per S)}  \\
		\hline
9.95 	&	0.55	&	6.60 	\\
10.87 	&	0.56	&	1.18 	\\
12.40 	&	0.81	&	3.24 	\\
13.81 	&	0.36	&	0.15 	\\
15.35 	&	2.26	&	1.00 	\\
17.55 	&	0.96	&	0.98 	\\
19.45 	&	2.06	&	2.58 	\\
21.35* 	&	1.86	&	1.18 	\\
25.93 	&	1.96	&	2.63 	\\
		\bottomrule[1pt]%
		{*Gaussian Component}
		\end{tabular}
	\end{table}
%========Table 5==============

\end{document}